\documentclass[a4paper]{jpconf}
\usepackage{graphicx}
\usepackage{cite}
\usepackage{amsmath}
\usepackage{amssymb}

\newcommand{\rr}{\textbf{r}}
\newcommand{\kk}{\textbf{k}}

\begin{document}
\title{Spectra of protons and nuclei in the energy range of ${10^{10}\div10^{20}}$~eV in the framework of Galactic cosmic ray origin}

\author{Nikolay~Volkov, Anatoly~Lagutin, Alexander~Tyumentsev, Roman~Raikin}

\address{Altai State University, Radiophysics and Theoretical Physics Department. 61 Lenin ave., Barnaul, 656049, Russia.}

\ead{volkov@theory.asu.ru, lagutin@theory.asu.ru, tyumentsev@theory.asu.ru, raikin@theory.asu.ru}

\begin{abstract}
We consider the problem of the cosmic ray spectrum formation assuming that
cosmic rays are produced by Galactic sources. The anomalous diffusion equation
proposed in our recent papers is used to describe cosmic ray propagation
in the interstellar medium. We show that in the framework of this approach and with
generation spectrum exponent $\gamma=2.85$ it is possible to reproduce locally observed
basic features of cosmic rays in the energy region of $10^{10}\div 10^{20}$~eV:
difference between spectral exponents of protons and other nuclei, mass composition
variation, ``knee'' problem, flattening of the primary spectrum at $E \geq 10^{18}\div 10^{19}$~eV.
The crucial model predictions for the mass composition behaviour in the ultra-high energy region are discussed.
\end{abstract}

\section{Introduction}

The problem of ultra-high energy cosmic ray (UHECR) origin remains unsolved, despite significant progress made in recent years. The suppression of the cosmic ray flux at energy of $(4\div 6)\times 10^{19}$~eV being established unambiguously can not be reliably identified as GZK-cutoff due to the apparent differences between Telescope Array/HiRes and Auger mass composition results~\cite{AugerC, TAC}. As the Auger data indicated a nuclear UHECR composition becoming heavier with energy in the region of $4\div 40$~EeV, different models assuming that the UHECR spectra cutoff is not an extragalactic GZK feature are considered~\cite{Sc1, Sc2, Sc3}.

In this paper we examine the possibility for self-consistent description of all the basic features of the observed cosmic ray spectrum in the energy range of $10^{10}\div 10^{20}$~eV within the Galactic origin scenario. We assume the existence of Galactic sources that accelerate particles up to $\sim 10^{20}$~eV and take into account highly inhomogeneous (fractal-like) distribution of matter and magnetic fields in the Galaxy that leads to extremely large free paths of particles (``Levy flights''), along with an overwhelming contribution to the cosmic ray fluxes observed above $\sim 10^{18}$~eV from particles reaching the Solar System without scattering. The crucial model predictions for the mass composition behaviour in the ultra-high energy region are presented.

\section{Anomalous diffusion model}

In our recent papers an anomalous diffusion (AD) model for solution of the ``knee'' problem in primary cosmic-rays spectrum was proposed~\cite{L1, L2, E1, L3}. The ``anomaly'' in this model results from the presence of extremely large free paths (Levy flights) of particles between Galactic inhomogeneities and also the long-lasting trapping of particles in very strong magnetic field regions (Levy traps).
Thus the following asymptotic behaviour of the probabilities of free paths ($p(x,E)$) and time, during which particles are caught in magnetic traps ($q(\tau,E)$), takes place:

\begin{equation}
\int\limits_{|x| > r} p(x,E) d x \propto A(E,\alpha)r^{-\alpha}, \quad 0 < \alpha < 2;
\quad\int\limits_t^{\infty} q(\tau,E) d \tau \propto B(E,\beta)t^{-\beta},\quad 0 < \beta < 1.
\end{equation}

The cosmic rays propagation in the fractal-like interstellar Galactic medium without energy losses and nuclear interactions can be described by the anomalous diffusion equation. The equation for the concentration of particles with an energy $E$, generated in a fractal medium by Galactic sources with a distribution density $S(\rr,t,E)$ can be written as
\begin{equation}~\label{SuperEq}
\frac{\partial N}{\partial t}=-D(E,\alpha,\beta)\mathrm{D}_{0+}^{1-\beta}(-\Delta)^{\alpha/2} N(\rr,t,E) + S(\rr,t,E).
\end{equation}
Here $\mathrm{D}_{0+}^{\mu}$ denotes the Riemann-Liouville fractional derivative~\cite{S1} and $(-\Delta)^{\alpha/2}$ is the fractional Laplacian (`Riesz operator')~\cite{S1}. The anomalous diffusivity $D(E,\alpha,\beta) \sim A(E,\alpha)/B(E,\beta) = D_0(\alpha,\beta)E^{\delta}$.

The Green's function $G(\rr,t,E;E_0)$, corresponding to~\eqref{SuperEq}, can be found from the equation  
\begin{equation}~\label{SuperEqG}
\frac{\partial G(\rr,t,E;E_0)}{\partial t}=-D(E,\alpha,\beta)\mathrm{D}_{0+}^{1-\beta}(-\Delta)^{\alpha/2} G(\rr,t,E;E_0) +\delta({\rr})\delta(t)\delta(E-E_0).
\end{equation}

With the use of the Laplace-Fourier transformations and formulae~\cite{S1}
$$
\int\limits_0^\infty e^{-\lambda t}\mathrm{D}_{0+}^{\beta} G(\rr,t,E;E_0) dt = \lambda^{\beta} \int\limits_0^\infty e^{-\lambda t} G(\rr,t,E;E_0) dt = \lambda^{\beta} \tilde{G}(\rr,\lambda,E;E_0),
$$
$$
\int\limits_{{\rm R}^3} \mathrm{e}^{ik \rr} (-\Delta)^{\alpha/2} G(\rr,t,E;E_0) d \rr = |k|^{\alpha}\int\limits_{{\rm R}^3} \mathrm{e}^{ik \rr} G(\rr,t,E;E_0) d \rr = |\kk|^{\alpha} \tilde{G}(\kk,t,E;E_0)
$$
we obtain
$$\tilde{G}(\kk,\lambda,E;E_0)=\delta(E-E_0)\lambda^{\beta-1}\int\limits_0^\infty\exp\left(-\left[\lambda^{\beta}+D(E,\alpha,\beta)|\kk|^{\alpha}\right]y\right)dy.$$

\begin{figure}[bt]
\begin{center}
\includegraphics[width=0.75\textwidth]{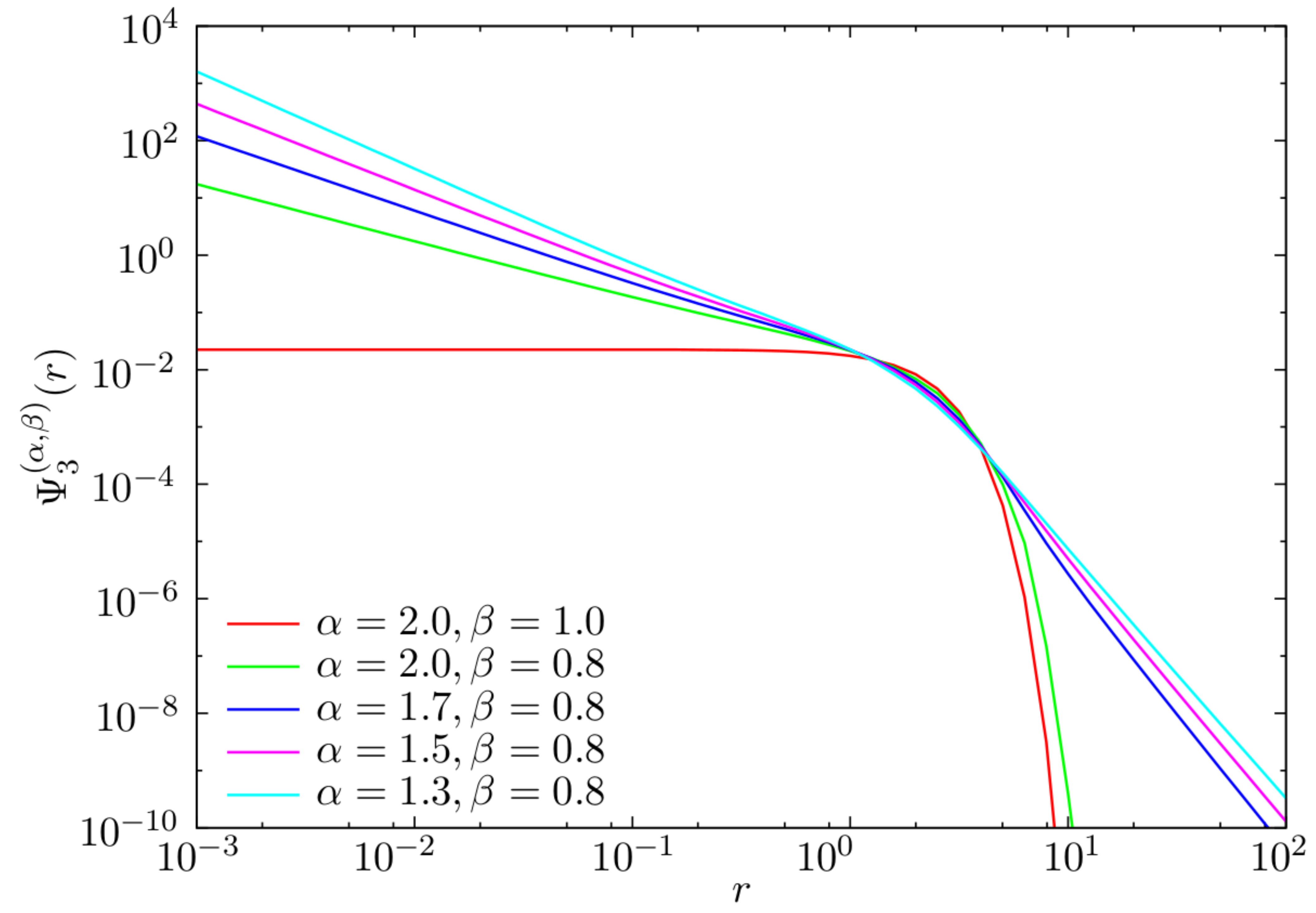}
\end{center}
\caption{Fractional stable distributions densities $\Psi_3^{(\alpha, \beta)}(r)$ }\label{fpsi}
\end{figure}

Then the inverse transform yields
\begin{equation}\label{eq:green}
G(\rr,t,E;E_0) = \delta(E-E_0)\left(D(E_0,\alpha,\beta)t^{\beta}\right)^{-3/\alpha}\Psi_3^{(\alpha, \beta)}\left(|\rr|\left(D(E_0,\alpha,\beta)t^{\beta}\right)^{-1/\alpha}\right),
\end{equation}
where
\begin{equation*}\label{eq:psi}
\Psi_3^{(\alpha, \beta)}(r)=\int\limits_0^\infty g_3^{(\alpha)}\left(r\tau^{\beta/\alpha}\right) g_1^{(\beta,1)}(\tau) \tau^{3\beta/\alpha}d\tau
\end{equation*}
is the density of fractional stable distribution~\cite{U1}, $g_3^{(\alpha)}(r)$ $(\alpha\leq 2)$ the three-dimensional spherically-symmetrical stable distribution and $g_1^{(\beta,1)}(t)$ is the one-sided stable distribution with characteristic exponent $\beta\leq 1$~\cite{U1}. The probability densities of $\Psi_3^{(\alpha, \beta)}(r)$ are shown in figure~\ref{fpsi}.

Using the Green's function~\eqref{eq:green}, we can write a solution of the anomalous diffusion equation for a point pulse source { $S(\rr,t,E)=S_{\text{im}} E^{-\gamma}\exp(-E/E_0(z))\delta(\rr)\Theta(T-t)\Theta(t)$} with emission time $T$ and an exponential cutoff at energy $E_0$
\begin{multline*}
N(\rr,t,E)=\frac{S_{\text{im}} E^{-\gamma}}{D(E,\alpha,\beta)^{3/\alpha}} \times\\
\times\left[\int\limits_{\max[0,t-T]}^{t}d\tau \tau^{-3\beta/\alpha}\Psi_3^{(\alpha,\beta)}\left(|\rr|(D(E,\alpha,\beta)\tau^{\beta})^{-1/\alpha}\right)\right]\exp\left(-\frac{E}{E_0(z)}\right).
\end{multline*}

\section{Energy spectrum of cosmic rays}

Types of sources and their distribution in space suggest their separation into three parts as follows.
$$
J(\rr,t,E) = J_G(\rr,E)+ J_L(\rr,t,E)+J_{NS}(\rr,E)
$$

Here
\begin{itemize}
\item $J_G$ is the global spectrum component determined by the multiple old ($t\geq 10^6$~yr) distant ($r\geq 1$~kpc) sources;
\item $J_L$ is the local component, i.e. the contribution nearby ($r< 1$~kpc) young ($t< 10^6$~yr) sources;
\item $J_{NS}$ is the flux of non-scattered particles.
\end{itemize}

\begin{figure}[htb]
\begin{center}
\includegraphics[width=\textwidth]{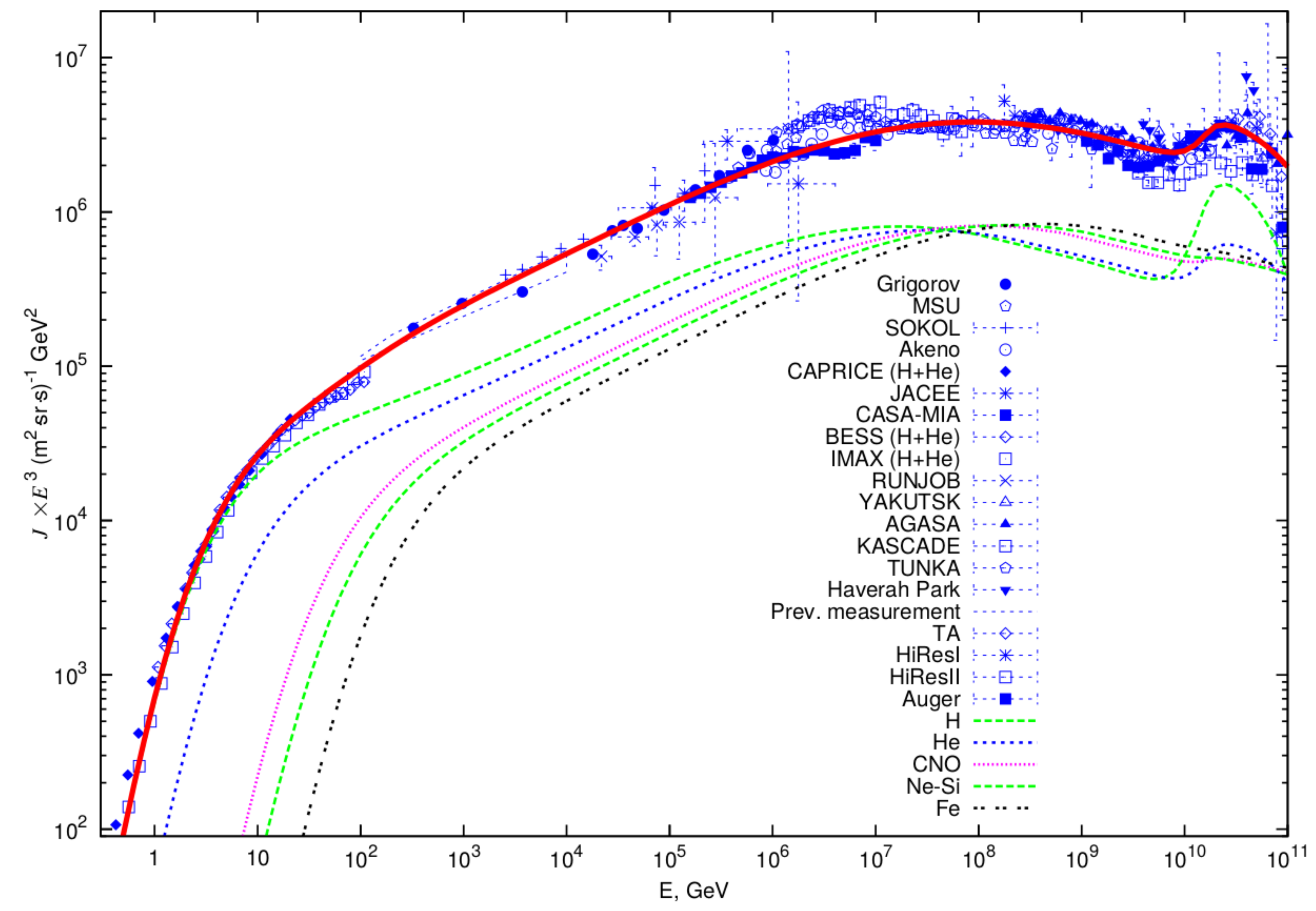}
\end{center}
\caption{Comparison of our calculations of primary cosmic ray spectra with experimental data in wide energy range of $10^{10}\div 10^{20}$~eV. Curves marked 'H', 'He', 'CNO', 'Ne-Si', 'Fe' and 'Total' represent our results.}\label{fsp1}
\end{figure}

\begin{figure}[htb]
\begin{center}
\includegraphics[width=.75\textwidth]{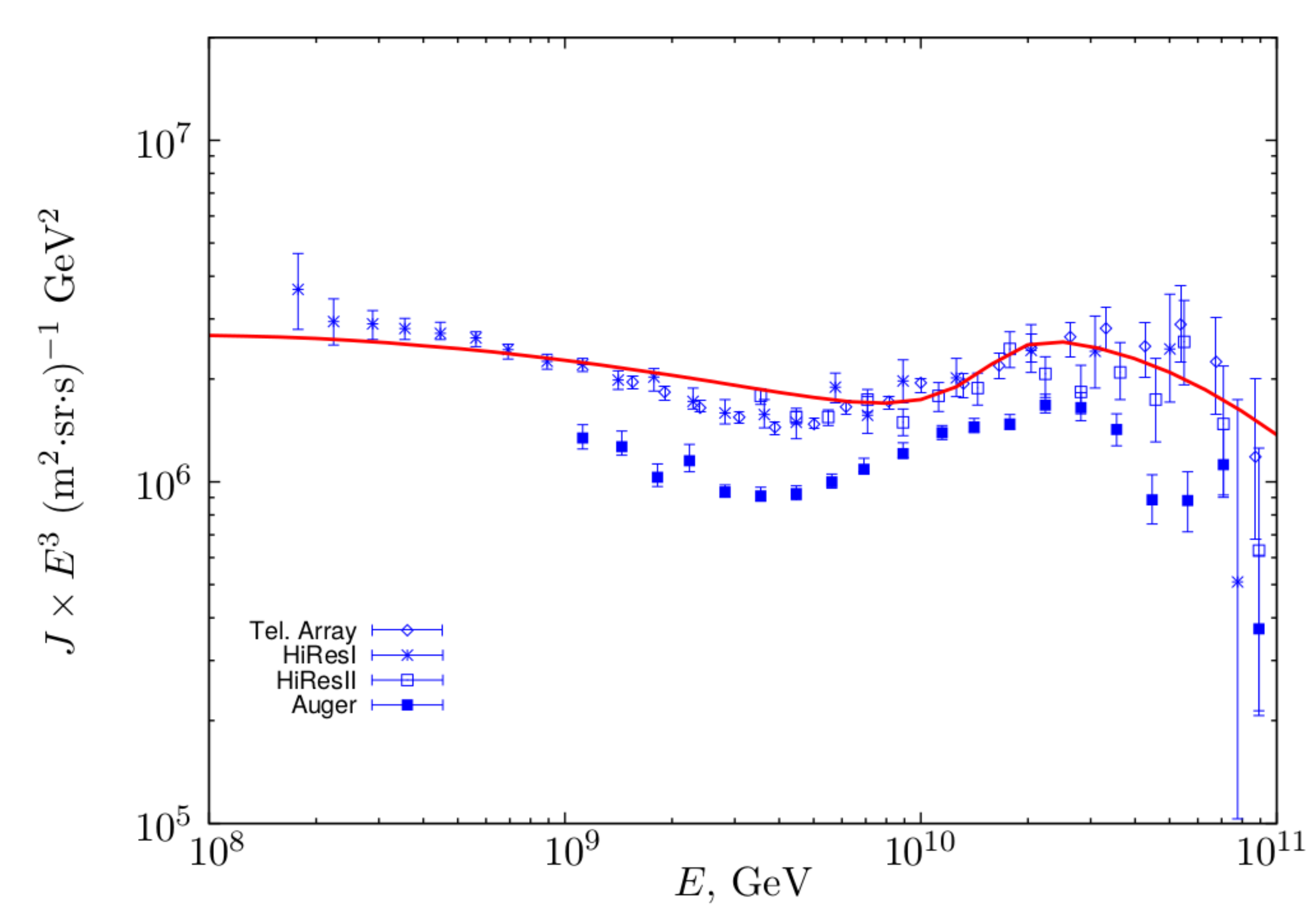}
\end{center}
\caption{Comparison of our calculations of primary cosmic ray spectra with experimental data of Telescope Array/HiRes~\cite{TelArray} and Auger~\cite{Auger} at ultrahigh energies.}\label{fsp2}
\end{figure}

The flux of non-scattered particles $J_{NS}$ is determined by the injected flux
$$
S_0E^{-\gamma}\exp(-E/E_0(z))
$$
and the Levy flight probability $p(E,>r)$. Taking into account that for a particle with energy $E$ the probability $p(E,>r)\sim A(E,\alpha)r^{-\alpha}$, $A(E,\alpha)\sim E^{\delta_L}$, we have
$$
J_{NS}(r,E) = S_{NS} E^{-\gamma+\delta_L}\exp(-E/E_0(z))r^{-\alpha}.
$$
We assume that this component is also formed by nearby ($r< 1$~kpc) sources, defining the spectrum in ultrahigh energy region, and provides the observed flattening of the spectrum at $E \geq 10^{18}$~eV.

Finally, for the global stationary component formed by old distant sources and the local component from the nearest sources the expression obtained by the solution of anomalous diffusion equation is as follows.
\begin{multline}\label{eq:sol}
	J(\rr,t,E) = \frac{v}{4\pi}\Biggl[S_{\text{s}}E^{-\gamma-\delta/\beta}+ \frac{S_{\text{im}} E^{-\gamma}}{D(E,\alpha,\beta)^{3/\alpha}}\times\Biggr.\\
	 \times\sum\limits_{\substack{r_j < 1\;\text{kpc}\\t_j < 10^6\;\text{yr}}} \int\limits_{\max[0,t_j-T]}^{t_j}d\tau
	\tau^{-3\beta/\alpha}\Psi_3^{(\alpha,\beta)}\left(|\rr_j|(D(E,\alpha,\beta)\tau^{\beta})^{-1/\alpha}\right)+\\
	+\Biggl. S_{NS}\sum\limits_{r_j < 1\;\text{kpc}}E^{-\gamma+\delta_L}|\rr_j|^{-\alpha}\Biggr]\exp\left(-\frac{E}{E_0(z)}\right).
      \end{multline}
In~\eqref{eq:sol} the first term is the contribution from the multiple old distant sources, the second term is the contribution from nearby young sources and the third term is the flux of non-scattered particles. To calculate the energy spectrum from nearby young sources, simulation of the Poisson ensemble of sources was carried out~\cite{Lagutin:2013}. The Poisson distribution parameter (average number of sources in the local region) was chosen $\sim10$. This estimation corresponds to number of the well-known nearest supernova remnants and pulsars with $t< 10^6$~yr. Coordinates and times of birth of the sources were generated randomly and uniformly in the space region $r<10^3$~pc and in the time interval $10^4\leq t< 10^6$~yr.

The parameters of the model are $\gamma = 2.85$, $\delta = 0.27$ as it was established earlier~\cite{L1}, $\delta_L = \delta/2$. The cut-off energy $E_0(z)=4\times 10^{10}z~\text{GeV}$ was fixed on the basis of comparisons with experimental data. To assess parameters $\alpha = 1.4$, $\beta = 0.8$, which are specific for our model, the results of studies of particles diffusion in cosmic and laboratory plasma have been used~\cite{Greco:2003,Perri:2008}. Finally, for the anomalous diffusivity we get the value of $D_0 \approx 10^{-4} \text{pc}^{1.4}/\text{yr}^{0.8}$~\cite{L3}. The emission time $T=10^4$~yr.

Our results of calculations of energy spectra of cosmic rays are shown in figures~\ref{fsp1}, \ref{fsp2}. Note, that the ``fine structure'' of spectrum around the knee can be described satisfactorily in the framework of the model with two types of sources (see~\cite{Ltt}).

\begin{figure}[bt]
\begin{center}
\includegraphics[width=.95\textwidth]{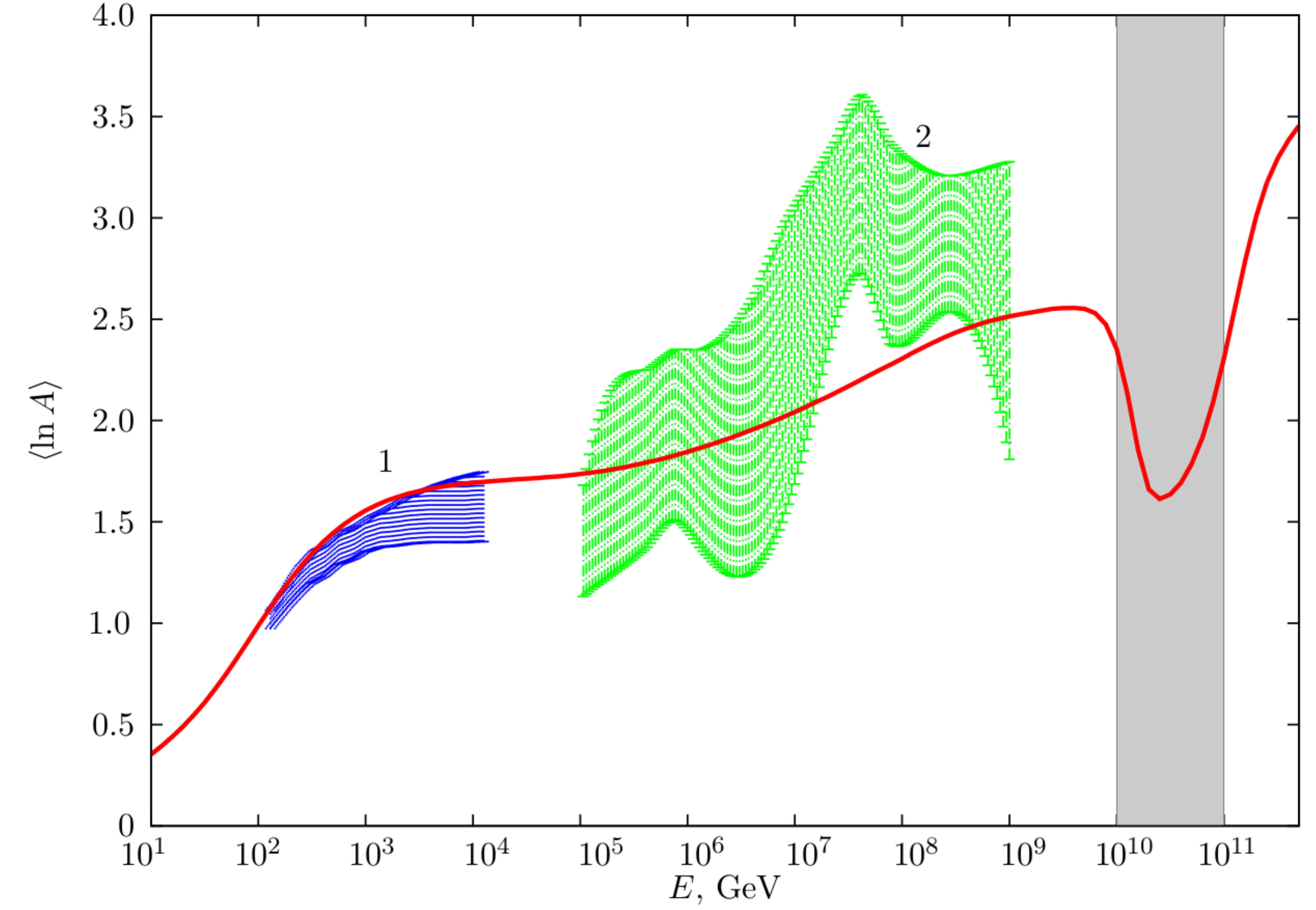}
\end{center}
\caption{Mean logarithmic cosmic ray mass vs. primary particle energy.
Dashed areas --- experimental data~\cite{S2},\cite{H1}. }\label{fsp3}
\end{figure}

An average mass number variation with energy according to our model is shown in figure~\ref{fsp3}. It was established that in the region of $(2\div 5)\times 10^{19}$~eV the fraction of protons is $\sim 40\%$, while the fraction of iron is $\sim 15\%$. A key feature of the model is the essential dip in mean logarithm of the mass number in the region of $(4\div 25)\times 10^{18}$~eV (from 2.6 to 1.6). The weighting of the mass composition is predicted at $E> 2.5\times 10^{19}$~eV and $\langle\ln A\rangle$ reaches $\sim 3$ at $E=2\cdot10^{20}$~eV. The presence of such a dip according to the experimental data could be construed in favour of our model. Note, that recent results of Auger Collaboration exhibit similar feature in $\langle\ln A\rangle$ variation with energy, although at an order of magnitude lower energies~\cite{pao}.

\section{Conclusions}

Assuming the Galactic origin of cosmic rays we propose a mechanism for the formation of ultra-high energy protons and nuclei spectra observed in the Solar System, that reproduces all the basic features of cosmic ray spectrum observed experimentally.

It is shown that taking into account both the diffusion contribution from nearby young and old distant sources and also the contribution of non-scattered radiation, due to the presence of large free paths (Levy flights) caused by anomalous diffusion, a description of the experimental data in the energy range $10^{10}\div10^{20}$~eV can be obtained if we assume that the particle injection spectrum in the Galactic sources is
$$
S=S_0E^{-2.85}\exp(-E/E_0(z)),\quad E_0(z)=4\times 10^{10}z~\text{GeV}.
$$

The following basic model predictions are expressed:
\begin{itemize}
\item the steepening of all particles spectrum in the range $E>5\times 10^{19}$~eV is due to the proton flux cut-off caused by energy limitation of the Galactic particle accelerators;
\item at the energies $2\div 5\times 10^{19}$~eV the fraction of protons is $\sim 40$~\%, and the iron fraction is about $15$~\% (normal composition);
\item in the range $4\div 25\times10^{18}$~eV mean logarithmic cosmic ray mass $\langle\ln A\rangle$ decreases rapidly from $2.6$ to $\langle\ln A\rangle_{\min}\sim 1.6$;
\item considerable weighting of the mass composition is observed at $E>2.5\cdot10^{19}$~eV; for $E=2\cdot10^{20}$~eV mean logarithmic cosmic ray mass is $\langle\ln A\rangle\sim 3$.
\end{itemize}

\section*{Acknowledgement}
This work was suppored by the Russian Foundation for Basic Research grant No. 14-02-31524. The authors are greatly indebted to anonymous referee for valuable comments.


\section*{References}
\begin{thebibliography}{10}

\bibitem{AugerC} Abbasi R U et al 2008 {\it Phys. Rev. Lett.} {\bf 100} 101101

\bibitem{TAC} Abraham J et al 2008 {\it Phys. Rev. Lett.} {\bf 101} 061101

\bibitem{Sc1} Aloisio R, Berezinsky V, Gazizov A 2011 {\it Astropart. Phys.} {\bf 34} 620

\bibitem{Sc2} Calvez A, Kusenko A, Nagataki S 2010 {\it Phys. Rev. Lett.} {\bf 105} 091101

\bibitem{Sc3} Fang K, Kotera K, Olinto A V 2013 {\it JCAP} {\bf 03} 010.

\bibitem{L1} Lagutin A~A, Nikulin Yu~A, Uchaikin V~V 2001 {\em Nucl. Phys. B} {\bf 97} 267

\bibitem{L2} Lagutin A~A, Uchaikin V~V  2003 {\em NIM B\/} {\bf B201} 212

\bibitem{E1} Erlykin A~D, Lagutin A~A, Wolfendale A~W 2003 {\em  Astropart. Phys.} {\bf 19} 351

\bibitem{L3} Lagutin A~A and Tyumentsev A~G 2004 {\em Izv. Altai. Gos. Univ.\/} {\bf 35} 4 (in Russian)

\bibitem{S1} Samco S~G, Kilbas A~A and Marichev O~I 1993 {\em Fractional integrals and derivatives --- Theory and Applications\/} (New York: Gordon and Breach) p 976

\bibitem{U1} Uchaikin V~V and Zolotarev V~M 1999 {\em Chance and Stability\/} (Netherlands,  Utrecht: VSP) p 594

\bibitem{TelArray} Abu-Zayyad T, Aida R, Allen M {(Telescope Array Experiment)} 2013 {\em Astrophys. Journ. Lett.} {\bf 768:L1} 1

\bibitem{Auger} Abraham J, Abreu P, Aglietta M {(Pierre Auger Collaboration)} 2010 {\em Phys. Lett. B} {\bf 685} 239

\bibitem{Lagutin:2013} Lagutin A~A, Volkov N~V and Tyumentsev A~G 2013 {\em Bulletin of the Russian Academy of Sciences: Physics} {\bf 77} 1312 

\bibitem{Greco:2003} Greco A, Taktakishvili A~L, Zimbardo G and et~al 2003 {\em J. Geophys. Res.\/} {\bf 108} 1

\bibitem{Perri:2008} Perri S and Zimbardo G 2008 {\em J. Geophys. Res.\/} {\bf 113}

\bibitem{Ltt} Lagutin A~A, Tyumentsev A~G, Yushkov A V  2008 {\em Nucl. Phys. B (Proc. Suppl.)} {\bf 175-176} 555

\bibitem{pao} Aab A, Abreu P, Aglietta M {(Pierre Auger Collaboration)} 2014 {\em Phys. Rev. D.\/} {\bf 90} 122005

\bibitem{S2} Shibata T 1999 {\em Nucl. Phys. B.\/} {\bf 75A} 22

\bibitem{H1} H{\" o}randel J~R 2003 {\em Astropart. Phys.\/} {\bf 19} 193

\end{thebibliography}
\end{document}